\documentclass[12pt,aps,prb,preprint]{revtex4}   

\usepackage{amsmath}    
\usepackage{graphicx}   

\begin{document}

\title{An alternative formulation of the\\ magnetostatic boundary value problem}
\author{L. Nasser}
 \affiliation{Columbia College, Department of Science and Mathematics, Chicago,
IL 60605}
 \email{lnasser@colum.edu}  
\author{Z. Chacko}
\affiliation{University Of Maryland, Department of Physics, College Park,
MD 49007}
\date{\today}

\begin{abstract} We present an alternative formulation of the magnetostatic boundary value problem which is useful for calculating the magnetic field around a magnetic material placed in the vicinity of steady currents. The formulation differs from the standard approach in that a single-valued scalar potential, plus a vector field that depends on the given currents but not on the magnetic material, are used to obtain the total magnetic field instead of a magnetic vector potential. We illustrate the method by a few sample computations. 
\end{abstract}

\maketitle

\section{Introduction}Solving for the magnetic field around linear, isotropic magnetic material placed in the vicinity of steady current-carrying conductors is one of the most important uses of magnetostatics. The evaluation of the fields follows from a standard manipulation of Maxwell's equations, and is treated extensively in the literature.\cite{stratton}$^{,}$\cite{jackson}$^{,}$\cite{poly}$^{,}$\cite{jefi}
The following description is limited to linear, isotropic materials for simplicity though the method itself is more general. The relevant Maxwell's equations are:
\begin{eqnarray}
\label{h}
\nabla\times\vec{H} & = & \frac{4\pi}{c}\vec{J}, \\
\label{b}
\nabla\cdot\vec{B} & = & 0, 
\end{eqnarray}
where $\vec{J}$ is a given surface current and we impose continuity of the normal component of $\vec{B}$ and the tangential component of $\vec{H}$ across the surface as boundary conditions:
\begin{eqnarray}
\vec{B}_{\textrm{int}}\cdot\hat{n}|_{s} & = & \vec{B}_{\textrm{ext}}\cdot\hat{n}|_{s} \\
\hat{n}\times\vec{H}_{\textrm{int}}|_{s} & = &  \hat{n}\times\vec{H}_{\textrm{ext}}|_{s}
\end{eqnarray} 
where $\hat{n}$ is the unit normal at the surface. \\
In the standard procedure for solving this problem we define the vector potential by
\begin{equation}
\label{veca}
\nabla\times\vec{A}=\vec{B}
\end{equation}
which automatically satisfies equation (\ref{b}). For convenience, we choose $\nabla\cdot\vec{A}=0$ and then equation (\ref{h}) yields
\begin{eqnarray}
\label{nabs}
\left(\nabla\times\frac{1}{\mu}\left(\nabla\times\vec{A}\right)\right)&=&\frac{4\pi}{c}\vec{J},\\
\label{A}
\nabla^{2}\vec{A}&=&-\frac{4\mu\pi}{c}\vec{J},
\end{eqnarray}
where $\mu=1$ outside the magnetic material.

In a two-dimensional problem a single component of $\vec{A}$ will often suffice to obtain the magnetic fields. However, in any truly three-dimensional problem solving equation (\ref{A}) to yield all three components of $\vec{A}$ can represent a formidable task. The impracticality of applying a standard magnetic scalar potential approach to such a problem arises from the fact that the scalar potential $\phi$, defined in a current-free region as
\begin{equation}
\label{hscal}
\vec{H}=-\nabla\phi
\end{equation}
is not single valued. This can be seen if we rewrite equation (\ref{h}) in terms of the scalar potential using (\ref{hscal}) and integrate over an area transverse to and including all of the transverse current. Applying Stoke's theorem, we find
\begin{equation}
\label{}
\Delta\phi\big|_{\textrm{closed loop}}=\frac{4\pi}{c}I_{\textrm{enclosed}}.
\end{equation}
Hence the scalar potential picks up a contribution every time we go around a source current. This multiple-valuedness can become awkward, especially in situations where the magnetic body itself carries a free current. In this paper, we will discuss a formulation that overcomes the major shortcomings of the procedure we have just discussed. Essentially, the method consists of removing the rotational component of the magnetic field from the total magnetic field, and evaluating the remainder as the gradient of a single-valued scalar potential. It will be shown by means of some examples that this approach offers the simplest solution to any given, fully three-dimensional problem. It is important to note that this procedure is known to engineers\cite{zlo} but it appears to have eluded the notice of the general physics community. Moreover, their emphasis is in finite element methods while here we demonstrate its power in analytical calculations. there have been other publications dealing with alternate methods of calculating the magnetostatic field\cite{jefi2}, but to our knowledge, there isn't anything in the physics literature that uses the scalar potential method described here.

\section{The New Formulation} We begin by defining a new vector $\vec{K}$ as
\begin{eqnarray}
\label{kdiv}
\nabla\cdot\vec{K}& = & 0, \\
\label{kcurl}
\nabla\times\vec{K} & = & \frac{4\pi}{c}\vec{J},
\end{eqnarray}
from which it is clear that $\vec{K}$ is the magnetic intensity $\vec{H}$ in the absence of magnetic materials. We then have
\begin{equation}
\label{ }
\vec{K}(\vec{r})=\frac{4\pi}{c}\int\frac{\vec{J}\times(\vec{r}-\vec{r^{\prime}})}{|\vec{r}-\vec{r^{\prime}}|^{3}}d^{3}\vec{r}^{\prime}.
\end{equation}
Now define the vector $\vec{C}$ as
\begin{equation}
\label{ }
\vec{C}=\vec{H}-\vec{K}.
\end{equation}
The equations satisfied by $\vec{C}$ are 
\begin{eqnarray}
\label{cdiv}
\nabla\times\vec{C} & = & 0, \\
\nabla\cdot(\mu\vec{C})& = &-\nabla\cdot(\mu\vec{K}),  
\end{eqnarray}
since $\hat{n}\cdot\vec{K}$ is continuous across the surface by equation (\ref{kdiv}). From equation (\ref{cdiv}) it is clear that we may write $\vec{C}=-\nabla\phi$, where $\phi$ is a single-valued scalar potential.\cite{pl} This function is continuous everywhere but is not differentiable at the boundary surface. It follows that we must distinguish between the interior potential $\phi_{\textrm{int}}$ and the exterior potential $\phi_{\textrm{ext}}$, and solve Laplace's equation
\begin{eqnarray}
\nabla^{2}\phi_{\textrm{int}}& = & 0, \\
\nabla^{2}\phi_{\textrm{ext}}& = & 0,
\end{eqnarray}
subject to the following boundary conditions:
\begin{eqnarray}
\label{b1}
(\hat{n}\times\nabla\phi_{\textrm{int}})\big|_{\textrm{surface}} & = & (\nabla\times\phi_{\textrm{ext}})\big|_{\textrm{surface}}, \\
\label{b2}
(\mu-1)\vec{K}\cdot\hat{n}\big|_{\textrm{surface}} & = &(\mu\nabla\phi_{\textrm{int}}-\nabla\phi_{\textrm{ext}})\cdot\hat{n}\big|_{\textrm{surface}},
\end{eqnarray}
with $\nabla\phi\rightarrow0$ as $|\vec{r}|\rightarrow\infty$. Once the scalar potential is known, we obtain the magnetic field from
\begin{equation}
\label{H}
\vec{H}=-\nabla\phi+\vec{K}.
\end{equation}
Exact solutions to the equations above can sometimes be obtained by writing an expansion for $\phi$ in a basis appropriate to the geometry of the problem, and matching the expressions thus obtained for the interior and exterior regions at the boundary in accord with (\ref{b1}) and (\ref{b2}). A few such examples will be evaluated shortly. In addition, for more elaborate geometries which are less prone to yielding closed-form algebraic solutions, the scalar potential $\phi$ may be determined numerically, often with much greater ease than the evaluation of the corresponding vector potential. Clearly the computational economy of this method is one of its most striking advantages, but we must bear in mind its generality as well; the method may be applied to any problem, irrespective of whether the magnetic material carries a source current or not.

\section{Analogy with electrostatics} Not surprisingly, this procedure has a simple analogy in electrostatics\cite{zhou}$^{,}$\cite{lerner}. Consider a linear, isotropic dielectric in free space subject to an electric field $\vec{E_{0}}$ which may arise from charges embedded in the dielectric medium. The source charge distribution is assumed to remain unchanged, and the equations that determine the electric field are
\begin{eqnarray}
\nabla\cdot(\epsilon\vec{E}) & = & 0, \\
\nabla\times\vec{E} & = & 0, 
\end{eqnarray}
with boundary conditions
\begin{eqnarray}
\hat{n}\times\vec{E_{\textrm{int}}}\big|_{\textrm{surface}} & = & \hat{n}\times\vec{E_{\textrm{ext}}}\big|_{\textrm{surface}},  \\
\epsilon\vec{E}_{\textrm{int}}\cdot\hat{n}\big|_{\textrm{surface}} & = &  \vec{E}_{\textrm{ext}}\cdot\hat{n}\big|_{\textrm{surface}},
\end{eqnarray}
with $|\vec{E}|\rightarrow0$ as $|\vec{r}|\rightarrow0$. Writing
\begin{eqnarray}
\vec{E}_{\textrm{int}} & = & -\nabla\phi_{\textrm{int}}+\vec{E}_{0}, \\
\vec{E}_{\textrm{ext}} & = & -\nabla\phi_{\textrm{ext}}+\vec{E}_{0}, 
\end{eqnarray}
we get
\begin{eqnarray}
\nabla^{2}\phi_{\textrm{int}}& = & 0, \\
\nabla^{2}\phi_{\textrm{ext}}& = & 0,
\end{eqnarray}
with boundary conditions
\begin{eqnarray}
\label{e1}
(\hat{n}\times\nabla\phi_{\textrm{int}})\big|_{\textrm{surface}} & = & (\nabla\times\phi_{\textrm{ext}})\big|_{\textrm{surface}}, \\
\label{e2}
(\epsilon-1)\vec{E}_{0}\cdot\hat{n}\big|_{\textrm{surface}} & = &(\epsilon\nabla\phi_{\textrm{int}}-\nabla\phi_{\textrm{ext}})\cdot\hat{n}\big|_{\textrm{surface}},
\end{eqnarray}
with $\nabla\phi\rightarrow0$ as $|\vec{r}|\rightarrow\infty$. The correspondence between these equations and the previous set is clear: $(\mu-1)\vec{K}\cdot\hat{n}\big|_{\textrm{surface}} $ replaces $(\epsilon-1)\vec{E}_{0}\cdot\hat{n}\big|_{\textrm{surface}} $ as the ``source term''. Some examples of the use of this method are now presented.

\section{Examples of the Method}
\subsection{Infinitely long wire with current $I$ placed parallel to the axis of an infinitely long cylinder of permeability $\mu$} 
\begin{figure}
\begin{center}
\includegraphics[width=4in]{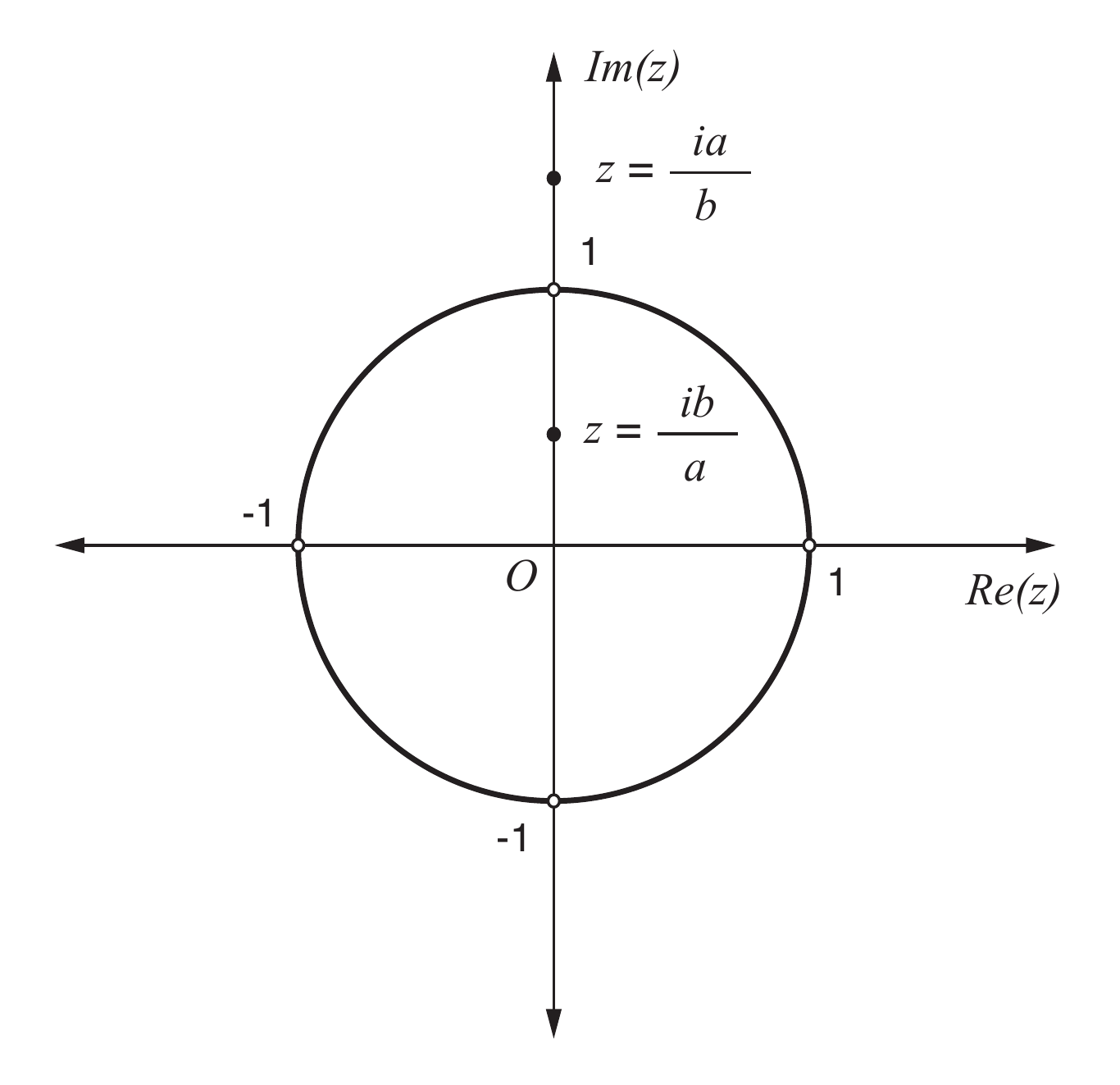}
\caption{Simple poles of $I_{n}$.}
\label{poles}
\end{center}
\end{figure}

Let the radius of the cylinder be $\rho = a$, and let the distance between the wire and the cylinder be $d$, with $d>a$. We take a point on the axis of the cylinder as the origin of coordinates, and  note that the problem has symmetry along the cylinder's axis, which we choose to be the z-direction. Choosing the yz plane to contain the wire, we have in polar coordinates
\begin{equation}
\label{kvecto}
\vec{K}=\frac{\lambda_{0}}{\rho^{2}-2\rho d\sin\theta+d^{2}}\left[d\cos\theta\hat{\rho}+(\rho-d\sin\theta)\hat{\theta}\right],
\end{equation}
where $\lambda_{0}$ is the current per unit length in the wire.
Writing the general forms of $\phi_{int}(\rho,\theta)$ and $\phi_{ext}(\rho,\theta)$ corresponding to solutions of $\nabla^{2}\phi=0$ with no z-dependance as
\begin{eqnarray}
\phi_{int} (\rho,\theta)& = & \sum_{m=1}^{\infty}\rho^{m}\left(A_{m}\cos m\theta+B_{m}\sin m\theta\right), \\
\phi_{ext}(\rho,\theta) & = & \sum_{m=1}^{\infty}\rho^{-m}\left(C_{m}\cos m\theta+D_{m}\sin m\theta\right), 
\end{eqnarray}
the boundary conditions yield
\begin{eqnarray}
A_{m}a^{2m} & = & C_{m}, \\
B_{m}a^{2m} & = & D_{m},
\end{eqnarray}
and 
\begin{equation}
\label{ }
(\mu-1)\vec{K}\cdot\hat{\rho}=\sum_{m=1}^{\infty}m(\mu+1)a^{m-1}\left[A_{m}\cos m\theta+B_{m}\sin m\theta\right].
\end{equation}
To determine the coefficients $A_{m}$ we note from (\ref{kvecto}) that
\begin{equation}
\label{ }
\vec{K}\cdot\hat{\rho}=\frac{\lambda_{0}d\cos\theta}{\rho^{2}-2\rho d\sin\theta+d^{2}}
\end{equation}
and we therefore need to evaluate $I_{n}$, given by
\begin{equation}
\label{acoeff}
I_{n}=\int_{0}^{2\pi}\frac{\cos\theta\cos n\theta}{a^{2}+d^{2}-2ad\sin\theta}d\theta,
\end{equation}
where $n\geq 1$. We set $z=e^{i\theta}$, and switch to an integration around the unit circle $|z|=1$, giving
\begin{equation}
\label{ }
I_{n}= \operatorname{Re}\left\{-\frac{1}{2ad}\oint\frac{(z^{2}+1)z^{n-1}}{(z^{2}-1)-iz\left(\frac{a}{d}+\frac{d}{a}\right)}dz\right\}.
\end{equation}
The integrand has simple poles at $z=ia/d$ and $z=id/a$ as shown in figure (\ref{poles}).
 Since only $z=ia/d$ lies within the unit circle, picking up the residue from that pole we find
\begin{equation}
\label{ }
I_{n} =
\begin{cases}
    0,  & \text{for } n=2m,  \\
    (-1)^{m}\frac{\pi}{d^{2}}\left(\frac{a}{d}\right)^{2m}, & \text{for } n=2m+1.
\end{cases}
\end{equation}
To evaluate the $B_{m}$ coefficients, we need to evaluate
\begin{equation}
\label{acoeff}
J_{n}=\int_{0}^{2\pi}\frac{\cos\theta\sin n\theta}{a^{2}+d^{2}-2ad\sin\theta}d\theta,
\end{equation}
and following an analogous procedure, we find
\begin{equation}
\label{ }
J_{n} =
\begin{cases}
    0,  & \text{for } n=2m+1,  \\
    (-1)^{m+1}\frac{\pi}{d^{2}}\left(\frac{a}{d}\right)^{2m-1}, & \text{for } n=2m.
\end{cases}
\end{equation}
With these results, we can now obtain
\begin{eqnarray}
A_{2m} & = & 0, \\
A_{2m+1}&=&\frac{(-1)^{m}\Lambda}{d^{2m+1}(2m+1)},\\
B_{2m}&=&-\frac{(-1)^{m}\Lambda}{d^{2m}2m},\\
B_{2m+1}&=&0,
\end{eqnarray}
where $\Lambda=\lambda_{0}(\mu-1)/(\mu+1)$. We then have
\begin{equation}
\label{phint}
\phi_{int}(\rho, \theta)=\sum_{m=0}^{\infty}\Lambda\left(\frac{(-1)^{m}}{2m+1}\right)\frac{\cos(2m+1)\theta}{d^{2m+1}}\rho^{2m+1}
-\sum_{m=1}^{\infty}\Lambda\left(\frac{(-1)^{m}}{2m}\right)\frac{\sin2m\theta}{d^{2m}}\rho^{2m}.
\end{equation}
Setting $y=i\rho e^{i\theta}/d$, we can recast (\ref{phint}) as
\begin{eqnarray}
\phi_{int}(\rho,\theta) & = & \operatorname{Im}\left\{ \sum_{m=0}^{\infty}\frac{y^{2m+1}}{2m+1}
-\sum_{m=1}^{\infty}\frac{y^{2m}}{2m}\right\},\nonumber\\
 & = & \frac{\Lambda}{2}\arctan\left[\frac{2d\rho\cos\theta}{d^{2}-\rho^{2}}\right] +
 \frac{\Lambda}{2}\arctan\left[\frac{\rho^{2}\sin2\theta}{d^{2}+\rho^{2}\cos2\theta}\right] .
\end{eqnarray}
Similarly, we obtain
\begin{equation}
\label{ }
\phi_{ext}(\rho,\theta)= \frac{\Lambda}{2}\arctan\left[\frac{2a^{2}}{\rho d}\left(\frac{\cos\theta}{1-\frac{a^{4}}{\rho^{2}d^{2}}}\right)\right]
+ \frac{\Lambda}{2}\arctan\left[\frac{a^{4}}{\rho^{2} d^{2}}\left(\frac{\sin2\theta}{1+\frac{a^{4}\cos2\theta}{\rho^{2}d^{2}}}\right)\right]
\end{equation}
The full field may now be determined from
\begin{eqnarray}
\vec{B} & = & \mu\vec{H}\nonumber \\
 & = &\mu(\vec{K}-\nabla\phi). 
\end{eqnarray}

\subsection{Sphere of constant permeability $\mu$ concentric with a ring of uniform current $I$}
\begin{figure}[h!]
\begin{center}
\includegraphics[width=3in]{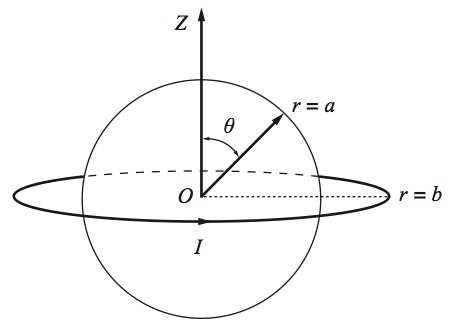}
\caption{Sphere of permeability $\mu$ and radius $a$ surrounded by a concentric ring of radius $b$ carrying a current $I$.}
\label{sphring}
\end{center}
\end{figure}

Consider a sphere of constant permeability $\mu$ and radius $a$, concentric with a circular ring of radius $b$, with $b>a$, that carries a current $I$, as shown in figure (\ref{sphring}). 
We wish to obtain the magnetic field everywhere in space. We begin by considering the case $b>a$. Using spherical coordinates, one can easily show\cite{arfken} that for $r<b$, the solution of equations (\ref{kdiv}) and (\ref{kcurl}) for the circular ring in free space is
\begin{eqnarray}
K_{r}& = & 2\lambda\sum_{n=0}^{\infty}f(n)r^{2n}P_{2n+1}(\cos\theta), \\
K_{\theta}& = & \lambda\sum_{n=0}^{\infty}g(n)r^{2n}P_{2n+1}^{1}(\cos\theta),
\end{eqnarray}
where $\lambda=\pi I/c$ and 
\begin{eqnarray}
f(n)& = & \frac{(-1)^{n}}{2^{n}b^{2n+1}}\frac{(2n+1)!!}{n!}, \\
g(n) & = & \frac{(-1)^{n}}{2^{n-1}b^{2n+1}}\frac{(2n-1)!!}{n!}, 
\end{eqnarray}
while for $r>b$ we have
\begin{eqnarray}
K_{r}& = & 2\lambda\sum_{n=0}^{\infty}\frac{a(n)}{r^{2n+3}}P_{2n+1}(\cos\theta), \\
K_{\theta}& = & \lambda\sum_{n=0}^{\infty}\frac{b(n)}{r^{2n+3}}P_{2n+1}^{1}(\cos\theta),
\end{eqnarray}
where
\begin{eqnarray}
\label{an}
a(n)& = & \frac{(-1)^{n}(2n+1)!!b^{2n+2}}{2^{n}{n!}}, \\
\label{bn}
b(n) & = &\frac{(-1)^{n}(2n+1)!!b^{2n+2}}{2^{n}{(n+1)!}}.
\end{eqnarray}
Now, since we are considering $b>a$,
\begin{eqnarray}
\label{muk}
(\mu-1)\vec{K}\cdot\hat{n}\big|_{\textrm{surface}} & = & (\mu-1)K_{r}\big|_{r=a}\nonumber \\
 & = & 2\lambda(\mu-1)\sum_{n=0}^{\infty}f(n)a^{2n}P_{2n+1}(\cos\theta).
\end{eqnarray}
We see that the generating term has no azimuthal dependence. We thus look for solutions of Laplace's equation with this symmetry:
\begin{eqnarray}
\Psi_{\textrm{int}} & = &\sum_{L}A_{L}r^{L}P_{L}(\cos\theta), \\
\Psi_{\textrm{out}} & = &\sum_{L}\frac{B_{L}}{r^{L+1}}P_{L}(\cos\theta).
\end{eqnarray}
From the boundary conditions (\ref{b1}) and (\ref{b2}) we obtain
\begin{eqnarray}
A_{L} & = & B_{L}a^{2L+1}, \\
\label{b3}
 (\mu-1)K_{r}\big|_{r=a}& = & \sum_{L}A_{L}a^{L-1}[1+L(1+\mu)]P_{L}(\cos\theta). 
\end{eqnarray}
By feeding (\ref{an}), (\ref{bn}) and (\ref{muk}) in (\ref{b3}) we finally obtain $A_{2n}=B_{2n}=0$, and
\begin{eqnarray}
A_{2n+1} & = & \frac{2\lambda(\mu-1)f(n)}{\mu(2n+1)+(2n+2)}, \\
B_{2n+1} & = & a^{4n+3}A_{2n+1}.
\end{eqnarray}
With the coefficients of the scalar potential expansion in hand, the field itself follows directly from equation (\ref{H}). 

\section{Conclusion and Further Work} 
\begin{figure}
\begin{center}
\includegraphics[width=3in]{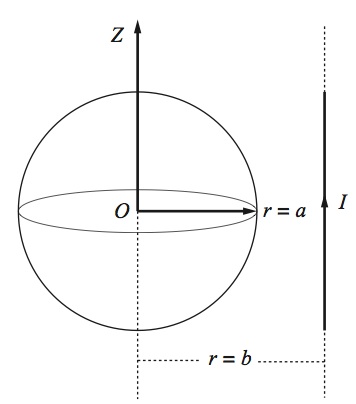}
\caption{Infinitely long, Straight wire carrying a current $I$ placed at a distance $d$ from the center of a sphere of permeability $\mu$ and radius $a$.}
\label{spherewire}
\end{center}
\end{figure}

We hope to have established that the procedure described in this paper offers a very general method of calculating the magnetic field when a source current is contained by, or placed in the vicinity of, a linear magnetic material. Moreover, this procedure is always simpler than the standard vector potential formulation for any truly three-dimensional problem. Applying the method requires nothing beyond the standard mathematical machinery needed to tackle boundary-value problems in electrostatics, and we hope that this alternative formulation will find its way into the standard curriculum on electricity and magnetism. There are many other fun problems students can solve to get practice with the method. For example, we suggest students could carry on where we left off and treat the case where $b<a$, and the ring of current is now embedded concentrically within the sphere. Another excellent example students could work on would be an infinitely long, straight wire with current $I$, placed a distance $d$ from the center of a sphere with permeability $\mu$ and radius $a$ as shown in figure (\ref{spherewire})
This problem can then be solved in the case where $a<d$ and then when the wire passes through the sphere for $a>d$.

\begin{acknowledgments}
We are grateful to J. R. Dorfman and J. L. Jim\'enez for their support and interest in this paper. Many thanks to Elizabeth Moss who prepared the figures, and to Andy Tillotson, Tim McCaskey and Ursula Perez-Salas for careful reading.
\end{acknowledgments}


\begin{thebibliography}{5}
\bibitem{pl} This separation of the field into two parts follows from an application of the Poincare lemma: We know that for any given scalar field $\psi$, $\nabla\times(\nabla\psi)=0$ always. The Poincare lemma then assures us that if a vector field $\vec{F}$ has the property that $\nabla\times\vec{F}=0$ in a given region of space, there exists a scalar function $\phi$ defined in that region such that $\vec{F}=\nabla\phi$.
\end{thebibliography}
\end{document}